\documentclass[12pt]{elsarticle}

\usepackage{graphics}
\usepackage{psfrag}
\usepackage{boxedminipage}

\usepackage{amssymb}

\biboptions{sort&compress}

\usepackage{color}
\usepackage{amsmath}
\usepackage{stmaryrd}

\newcommand{\Ti}{\mathcal{T}_i}
\newcommand{\ofxandt}{({\bf r}, t)}
\newcommand{\ofx}{({\bf r})}

\newcommand{\Ebold}{{\bf E}}
\newcommand{\Hbold}{{\bf H}}

\newcommand{\ebold}{{\bf e}}
\newcommand{\hbold}{{\bf h}}
\newcommand{\dd}{\text{d}}
\newcommand{\ileft}{_{\text{l}}}
\newcommand{\iright}{_{\text{r}}}
\newcommand{\whitem}{\textcolor{white}{-}}
\newcommand{\transp}{^{\text{T}}}

\newlength{\widebarargwidth}
\newlength{\widebarargheight}
\newlength{\widebarargdepth}
\DeclareRobustCommand{\widebar}[1]{%
  \settowidth{\widebarargwidth}{\ensuremath{#1}}%
  \settoheight{\widebarargheight}{\ensuremath{#1}}%
  \settodepth{\widebarargdepth}{\ensuremath{#1}}%
  \addtolength{\widebarargwidth}{-0.3\widebarargheight}%
  \addtolength{\widebarargwidth}{-0.3\widebarargdepth}%
  \makebox[0pt][l]{\hspace{0.3\widebarargheight}%
    \hspace{0.3\widebarargdepth}%
    \addtolength{\widebarargheight}{0.3ex}%
    \rule[\widebarargheight]{0.95\widebarargwidth}{0.1ex}}%
  {#1}}

\journal{Journal of Computational and Applied Mathematics}

\begin{document}

\begin{frontmatter}



\title{Efficient Large Scale Electromagnetics Simulations Using Dynamically Adapted Meshes with the Discontinuous Galerkin Method}


\author{Sascha M. Schnepp\fnref{affil}}
\ead{schnepp@gsc.tu-darmstadt.de}
\address{Graduate School of Computational Engineering, Technische Universitaet Darmstadt, Dolivostrasse~15, 64293~Darmstadt, Germany}
\fntext[affil]{The work of S.M.~Schnepp is supported by the 'Initiative for Excellence' of the German Federal
and State Governments and the Graduate School of Computational Engineering at Technische
Universitaet Darmstadt.}
\author{Thomas Weiland}
\ead{weiland@temf.tu-darmstadt.de}
\address{Institut fuer Theorie Elektromagnetischer Felder, Technische Universitaet Darmstadt, Schlossgartenstrasse 8, 64289 Darmstadt, Germany}

\begin{abstract}

A framework for performing dynamic mesh adaptation with the discontinuous Galerkin method (DGM) is presented.
Adaptations include modifications of the local mesh step size ($h$-adaptation) and the local
degree of the approximating polynomials ($p$-adaptation) as well as their combination.
The computation of the approximation within locally adapted elements is based on projections
between finite element spaces (FES), which are shown to preserve an upper limit of the electromagnetic energy. The formulation
supports high level hanging nodes and applies precomputation of surface integrals for increasing computational efficiency.
Error and smoothness estimates based on interface jumps are presented and applied to the
fully $hp$-adaptive simulation of two examples in one-dimensional space.
A full wave simulation of electromagnetic scattering form a radar reflector demonstrates the
applicability to large scale problems in three-dimensional space.

\end{abstract}


\end{frontmatter}


\section{Introduction}
\label{sec:intro}

The discontinuous Galerkin method~\cite{Reed1973,LeSaintRaviart1974} nowadays is a well-established method for solving
partial differential equations, especially for time-dependent problems.
It has been thoroughly investigated by Cockburn and Shu as well as Hesthaven and Warburton, who
summarized many of their findings in~\cite{Cockburn2001} and \cite{Hesthaven2008}, respectively.
Concerning Maxwell's equations in time-domain, the DGM has been studied in particular
in~\cite{Hesthaven2002Maxwell,Fezoui2005,Cohen2006,Gjonaj2006}. The former two apply
tetrahedral meshes, which provide flexibility for the generation of meshes also for complicated structures.
The latter two make use of hexahedral meshes, which allow for a computationally more efficient
implementation~\cite{Wirasaet:2010p1362}.

In~\cite{Cockburn2001} the authors state that
the method can easily deal with meshes with hanging nodes since no inter-element continuity is required, which
renders it particularly well suited for $hp$-adaptivity.
Indeed, many works are concerned with $h$-, $p$- or $hp$-adaptivity within the DG
framework. The first published work of this kind is presumably \cite{BeyOden:1996p1363},
where the authors consider linear scalar hyperbolic conservation laws in two space
dimensions. For a selection of other publications see 
\cite{Baumann:1999p179,Houston2001,Perugia:2002p172,Perugia:2003p181,Houston:2005p133} 
and references therein. The latter three are concerned with the adaptive
solution of Maxwell's equations in the time-harmonic case.

In this article, we are concerned with solving the Maxwell equations
for electromagnetic fields with arbitrary time dependence
in a three-dimensional domain $\Omega \subset \mathbb{R}^3$. They read
\begin{subequations}
\label{eq:maxwell}
\begin{eqnarray}
  \label{eq:Faraday}
  \nabla \times {\bf E} ({\bf r},t) &=& - \frac{\partial}{\partial t} {\bf B}({\bf r},t), \\
  \label{eq:Ampere}
  \nabla \times {\bf H} ({\bf r},t) &=& \textcolor{white}{-}\frac{\partial}{\partial t} {\bf D} ({\bf r},t) + {\bf J}({\bf r},t),
\end{eqnarray}
\end{subequations}
with the spatial variable ${\bf r} \in \Omega$ and the temporal variable $t$
subject to boundary conditions specified at the domain boundary $\partial\Omega$ and initial conditions specified at time $t_0$.
The vectors of the electric field and flux density are denoted by ${\bf E}$ and ${\bf D}$ and
the vectors of the magnetic field and flux density by ${\bf H}$ and ${\bf B}$.
The electric current density is denoted by ${\bf J}$. However, we assume the domain to be source free
and free of conductive currents (${\bf J} = {\bf 0}$).
Furthermore, we assume heterogeneous, linear, isotropic, non-dispersive and time-independent
materials in the constitutive relations
\begin{subequations}
\begin{eqnarray}
  \label{eq:BuH}
  {\bf B}({\bf r},t) & = & \mu({\bf r})\, {\bf H}({\bf r},t),\\
  \label{eq:DeE}
  {\bf D}({\bf r},t) & = & \epsilon({\bf r})\, {\bf E}({\bf r},t).
\end{eqnarray}
\end{subequations}
The material parameters $\mu$ and $\epsilon$ are the magnetic permeability and dielectric permittivity.
At the domain boundary, we apply either electric (${\bf n}\times {\bf E} = {\bf 0}$) or radiation boundary
conditions (${\bf n}\times {\bf E} = c\mu ({\bf n}\times {\bf n}\times {\bf H})$), where $c$ denotes the local
speed of light $c = (\epsilon \mu)^{-1/2}$.
We also introduce the electromagnetic energy $W$ contained in a volume $V$ obtained by
integrating the energy density $w$ as
\begin{equation}
  \label{eq:energyDensity}
  W(t) = \int_V w(t)\, \dd^3 {\bf r} 
  = \int_V \frac{1}{2} \left( \epsilon\ofx \Ebold\ofxandt^2 + \mu\ofx \Hbold\ofxandt^2 \right) \dd^3 {\bf r}.
\end{equation}

This paper focuses on a
general formulation of the DGM on non-regular hexahedral meshes as well as
the projection of solutions during mesh adaptation. The issues of optimality
of the projections and stability of the adaptive algorithm are addressed.
Special emphasis is put on discussing the computational efficiency. 
To the best of our knowledge, this is the first publication dealing with dynamical $hp$-meshes
for the Maxwell time-domain problem employing the DG method in three-dimensional space.

As they are key aspects of adaptive and specifically $hp$-adaptive methods,
we will also address the issues of local error and smoothness estimation.
This includes comments on the computational efficiency of the estimates.
As estimators are not at the core of this article the discussion is, however, rather short.

\section{Discontinuous Galerkin discretization on non-regular hexahedral grids}
\label{sec:discretization}

\subsection{Discretization of space}
\label{sec:discretization-space}

We perform a tesselation of the domain of interest $\Omega$ into $N$ hexahedra $\mathcal{T}_i$
such that the tesselation $\mathcal{T} = \bigcup_{i=1}^N \mathcal{T}_i$ is a polyhedral approximation of $\Omega$.
The tesselation is not required to be regular, however, it is assumed to be derivable from a regular root
tesselation $\mathcal{T}^0$ by means of element bisections. The number of element bisections along
each Cartesian coordinate, which is required to an obtain element $i$ of $\mathcal{T}$
is referred to as the refinement levels $L_{x,i}, L_{y,i}, L_{z,i}$. As we allow for anisotropic
bisecting the refinement levels of one element may differ. In case of isotropic refinement
we simply use $L_{i}$. The intersection of two neighboring elements
$\mathcal{T}_i \cap \mathcal{T}_k$ is called their interface, which we denote as $\mathcal{I}_{ik}$. As we consider non-regular
grids, every face $\mathcal{F}_j$ of a hexahedral element may
be partitioned into several interfaces depending on the number of neighbors $K$ such that
$\mathcal{F}_j = \bigcup_{k=1}^K \mathcal{I}_{ik}$. The face orientation
is described by the outward pointing unitary normal $\mathbf{n}_j$.
The union of all faces is denoted as $\mathcal{F}$, and the internal faces $\mathcal{F}\setminus\partial\Omega$
are denoted as $\mathcal{F}^{\mbox{\scriptsize{int}}}$.
Finally, the volume, area and length measures of elements, interfaces, faces and edges are referred
to as $|\mathcal{T}_i|$, $|\mathcal{I}_{ik}|$, $|\mathcal{F}_j|$ and $|\mathcal{T}_{d,i}|$, where $d$
denotes any of the Cartesian coordinates. Every element of the tesselation $\mathcal{T}$ is related
to a master element $\hat{\mathcal{T}} = [-1,1]^3$ through the mapping $G_i$
\begin{equation}
  \label{eq:Mapping}
  G_i :\, \hat{\mathcal{T}} \, \rightarrow \, \mathcal{T}_i : \, \hat{\bf r}\, \mapsto \, {\bf r} 
  = \left( 
    \frac{\hat{x}\,|\mathcal{T}_{x,i}|}{2} + x_{i,0}, 
    \frac{\hat{y}\,|\mathcal{T}_{y,i}|}{2} + y_{i,0}, 
    \frac{\hat{z}\,|\mathcal{T}_{z,i}|}{2} + z_{i,0}
  \right),
\end{equation}
where $d_{i,0}$ denotes the element center.

\subsection{General formulation}
\label{sec:general-formulation}

Multiplying Maxwell's equations~(\ref{eq:maxwell}) by
a test function $\psi\ofx  \in  H^1(\mathcal{T}_i)$, integrating over $\Ti$ and
performing integration by parts yields
\begin{subequations}
\label{eq:weakMaxwell}
  \begin{eqnarray}
    \label{eq:weakFaraday}
  \int\limits_{\Ti} \left(\psi \, \mu \frac{\partial}{\partial t}{\bf H}
    - (\mathbf{\nabla} \psi) \times {\bf E} \right) \dd^3 {\bf r} +
  \int\limits_{\partial \Ti} \psi \, (\mathbf{n} \times {\bf E})  \, \dd^2 {\bf r}  &=& 0, \\
  \label{eq:weakAmpere}
    \int\limits_{\Ti} \left( \psi \, \epsilon \frac{\partial}{\partial t}{\bf E} 
      + (\mathbf{\nabla} \psi )\times {\bf H} \right) \dd^3 {\bf r} - 
    \int\limits_{\partial \Ti} \psi \, (\mathbf{n} \times {\bf H})  \, \dd^2 {\bf r}  &=& 0,
  \end{eqnarray}
\end{subequations}
where the explicit dependencies of ${\bf r}$ and $t$ have been omitted. Equations~(\ref{eq:weakMaxwell})
constitute the generic weak formulation of the time-dependent Maxwell's equations. 
In the following, we will replace the exact
field solutions ${\bf E}$ and ${\bf H}$ by approximations using the discontinuous Galerkin framework.

The space and time continuous electromagnetic field quantities are 
approximated on $\mathcal{T}$ as
\begin{equation}
  \label{eq:approx}
    {\bf U}({\bf r},t) \approx
    {\bf U}_h ({\bf r},t) =
    \bigoplus_{i = 1}^N {\bf U}_i({\bf r},t),
\end{equation}
where ${\bf U} \in \{{\bf E}, {\bf H}\}$.
The element-local approximation ${\bf U}_i({\bf r},t)$ reads
\begin{equation}
  \label{eq:elocal}
  {\bf U}_i({\bf r},t)  =  \sum_{p} {\bf u}^{p}_i(t) \varphi^{p}_i({\bf r})
\end{equation}
with the polynomial basis functions $\varphi({\bf r})$ 
and the time-dependent vector of coefficients 
\begin{equation}
  \label{eq:elocalComponent}
  {\bf u}^{p}_i(t) = \left[u^{p}_{x,i}(t), u^{p}_{y,i}(t), u^{p}_{z,i}(t)\right]^\text{T}, 
\end{equation}
representing the numerical degrees of freedom.
The basis functions are defined with element-wise compact support, which
is an essential property of DG methods
\begin{equation}
  \label{eq:basisFcts}
  \varphi_i^p ({\bf r}) =
  \begin{cases}
    \varphi^p({\bf r}), &\mathbf{r} \in \mathcal{T}_i,\\
    0,                                & \text{otherwise}.
  \end{cases}
\end{equation}
We define the basis functions on the master element $\hat{\mathcal{T}}$
and obtain the element-specific basis through the mapping $G_i$ as
\begin{equation}
  \label{eq:basisMapping}
  \varphi_i = \hat{\varphi} \circ G_i^{-1}
\end{equation}
We employ Cartesian grids and tensor product basis functions of the form
\begin{equation}
  \label{eq:tensorBasis}
  \hat{\varphi}^p(\hat{\bf r}) = \bigotimes_{d\, \in \,\{x,y,z\}}\hat{\varphi}^{p_d}(\hat{r}_d),
\end{equation}
where $p$ is a multi-index obtained from all $p_d = 0..P_d$. We denote
by $P_i = (P_{x,i}, P_{y,i}, P_{z,i})$ the local maximum approximation orders
of element $\mathcal{T}_i$.
The finite element space (FES) $\mathcal{V}^P$ spanned by the basis functions
is given by the tensor product of the respective one-dimensional spaces
\begin{equation}
  \label{eq:DGtensorSpace}
  \mathcal{V}^P = \mathcal{V}_x^{P_x} \otimes \mathcal{V}_y^{P_y} \otimes \mathcal{V}_z^{P_z} \quad\text{with}\quad
  \mathcal{V}_d^{P_d} = \text{span}\{ \hat{\varphi}^{p_d }(\hat{r}_d);\, 0 \le p_d \le P_d \}
\end{equation}
The approximation may, thus,
make use of different orders $P_d$ in each of the coordinate directions, where
we drop the subscript if they are equal.
The basis functions are Legendre polynomials
scaled such that~\cite{Gjonaj2006}
\begin{equation}
  \label{eq:orthogonalBasis}
  \int_{\mathcal{T}_i} \varphi^p_i({\bf r})\varphi^q_i({\bf r})\, \dd^3 {\bf r} = 
\begin{cases}
|\mathcal{T}_i|, \quad &p = q\\
0, \quad &\text{otherwise}.
\end{cases}
\end{equation}
In the following the dependence of the spatial and temporal variable
is not written down explicitly.

If now we were to substitute the exact electromagnetic field solution $\{{\bf E}, {\bf H}\}$ for
its approximation the surface integral term of (\ref{eq:weakMaxwell}) cannot be evaluated straightforwardly
at the internal faces $\mathcal{F}^{\mbox{\scriptsize{int}}}$.
This is due to the ambiguity of the DG approximation at any interface
as a result of (\ref{eq:approx}) and (\ref{eq:basisFcts}). Weak continuity 
at internal faces is obtained locally by introducing numerical interface fluxes as
\begin{equation}
  \label{eq:fluxes}
  \int\limits_{\partial \Ti} \psi \, (\mathbf{n} \times {\bf U}^*)  \, \dd^2 {\bf r},
\end{equation}
where ${\bf U}^*$ is a unique interface value computed solely from ${\bf U}_i$ and ${\bf U}_k$, where
$\mathcal{T}_k$ is a neighboring element. Common choices include centered and upwind fluxes. The
centered interface value is given as
\begin{equation}
  \label{eq:cent}
    U^*_{d,ik} = \frac{1}{2}\big(  U_{d,k|\mathcal{I}_{ik}} + U_{d,i|\mathcal{I}_{ik}} \big).
\end{equation}
Computing the upwind value is more involved. It is obtained as the exact solution of Maxwell's
equations for piece-wise constant initial data after an infinitesimal time span, which is referred to as the Riemannian problem~\cite{LeVeque1990}.
For the $x$-component of the electric and magnetic field at an interface with normal ${\bf n}_z$ they read
\begin{subequations}
\label{eq:upw}
\begin{eqnarray}
  \label{eq:upwE}
    E_{x,ik}^* &=&  \frac{ ( Y_{k|\mathcal{I}_{ik}} E_{x,k|\mathcal{I}_{ik}} - H_{y,k|\mathcal{I}_{ik}} ) + ( Y_{i|\mathcal{I}_{ik}} E_{x,i|\mathcal{I}_{ik}} + H_{y,i|\mathcal{I}_{ik}})}{Y_{k|\mathcal{I}_{ik}} + Y_{i|\mathcal{I}_{ik}}}, \\
  \label{eq:upwH}
  H_{x,ik}^* &=& \frac{ ( Z_{k|\mathcal{I}_{ik}} H_{x,k|\mathcal{I}_{ik}} + E_{y,k|\mathcal{I}_{ik}} ) + ( Z_{i|\mathcal{I}_{ik}} H_{x,i|\mathcal{I}_{ik}} - E_{y,i|\mathcal{I}_{ik}})}
{Z_{k|\mathcal{I}_{ik}} + Z_{i|\mathcal{I}_{ik}}}.
\end{eqnarray}
\end{subequations}
with the intrinsic impedance and admittance
\begin{equation}
  \label{eq:FVZandY}
  Z = \sqrt{\frac{\epsilon}{\mu}}, \quad Y = \frac{1}{Z}.
\end{equation}
Other components are obtained by cycling the component indices and signs.

Note that centered fluxes preserve the Hamiltonian structure of Maxwell's equations while
this property does not carry over to the semi-discrete equations when applying upwind fluxes
due to the mixing of electric and magnetic quantities in~(\ref{eq:upw}).
Consequently, an energy conservation property~\cite{Fezoui2005,Gjonaj2006} can be obtained with the
centered flux formulation only, determining the kind of time integration schemes
to be used as well~\cite{Schnepp:2010JCP}. Our implementation
includes both flux types.

Having resolved the ambiguity at interfaces, we insert the approximations~(\ref{eq:approx})
into the weak formulation~(\ref{eq:weakMaxwell}) and follow the Galerkin procedure yielding
the semi-discrete DG formulation
\begin{subequations}
\label{eq:DGMaxwell}
  \begin{eqnarray}
    \label{eq:DGFaraday}
  \int\limits_{\Ti} \left(\varphi_i^{q_i} \, \mu \frac{\partial}{\partial t}{\bf H}_h
    - (\mathbf{\nabla} \varphi_i^{q_i}) \times {\bf E}_h \right) \dd^3 {\bf r} +
  \int\limits_{\partial \Ti} \varphi_i^{q_i} \, (\mathbf{n} \times {\bf E}_h^*)  \, \dd^2 {\bf r}  &=& 0, \\
  \label{eq:DGAmpere}
    \int\limits_{\Ti} \left( \varphi_i^{q_i} \, \epsilon \frac{\partial}{\partial t}{\bf E}_h 
      + (\mathbf{\nabla} \varphi_i^{q_i} )\times {\bf H}_h \right) \dd^3 {\bf r} - 
    \int\limits_{\partial \Ti} \varphi_i^{q_i} \, (\mathbf{n} \times {\bf H}_h^*)  \, \dd^2 {\bf r}  &=& 0,
  \end{eqnarray}
\end{subequations}
$\forall i = 1..N$, $\forall q_i = 0..P_i$. The volume integrals are referred to as the mass and stiffness terms,
the surface integrals represent face fluxes.
Note that no assumptions on the grid regularity have been made in the derivation.

\subsection{Employing non-regular grids containing high level hanging nodes}
\label{sec:empl-non-regul}

Due to the strictly element-local support of the basis and test functions, the DGM is highly suited
for the application on non-regular grids. The actual difference of the refinement levels $L_i$ and $L_k$
of neighboring elements, i.e., the level of hanging nodes, plays a minor role as shown in the
following.

Inspecting equations~(\ref{eq:DGMaxwell}) it is seen that the mass and stiffness terms are not affected by the
grid regularity as they are strictly local to the element $\mathcal{T}_i$. The flux term, however, involves neighboring elements
as well. Decomposing the surface integral into the six contributing face integrals
\begin{equation}
  \label{eq:surfaceFaces}
  \int\limits_{\partial \Ti} \varphi_i^{q_i} \, (\mathbf{n} \times {\bf U}_h^*)  \, \dd^2 {\bf r} = 
  \sum_{j=1}^6 \int\limits_{\mathcal{F}_{i,j}}\varphi_i^{q_i} \, ({\bf n}_j \times {\bf U}_h^*)  \, \dd^2 {\bf r},
\end{equation}
and considering centered fluxes for brevity each of these can be expressed as
\begin{equation}
  \label{eq:faceInterfaces}
  \frac{1}{2} \Bigg[\int\limits_{\mathcal{F}_{i,j}}\varphi_i^{q_i} \, ({\bf n}_j \times {\bf U}_{i})  \, \dd^2 {\bf r}
  + \sum\limits_k \int\limits_{\mathcal{I}_{ik|j}}\varphi_i^{q_i} \, ({\bf n}_j \times {\bf U}_k)  \, \dd^2 {\bf r}\Bigg].
\end{equation}
Accounting for the kind of non-regular grids described above, i.e.~grids obtained from a regular
root tesselation,  requires no more
than summing up the contributions of all neighboring elements to the total flux. This is
independent of the hanging node levels as well as the actual number of neighboring elements.
Inserting the approximation (\ref{eq:elocal}) into (\ref{eq:faceInterfaces}) yields
\begin{equation}
  \label{eq:faceInterfaceApprox}
  \frac{1}{2} \Bigg[  \sum_{p_i} {\bf n}_j \times {\bf u}^{p_i}_i \int\limits_{\mathcal{F}_{i,j}}\varphi_i^{q_i}\varphi^{p_i}_i \, \dd^2 {\bf r}
    + \sum_k \sum_{p_k} {\bf n}_j \times {\bf u}^{p_k}_k \int\limits_{\mathcal{I}_{ik|j}}\varphi_i^{q_i}\varphi^{p_k}_k \, \dd^2 {\bf r}
  \Bigg].
\end{equation}
Again, the first integral term does not depend on the grid regularity.
Assuming ${\bf n}_j$ to be aligned with the $z$-coordinate and to point towards positive direction it amounts to
\begin{equation}
  \label{eq:faceInterfaceRegular}
  \int\limits_{\mathcal{F}_{i,j}}\varphi_i^{q_i}\varphi^{p_i}_i \, \dd^2 {\bf r} = 
  \hat{\varphi}_{z}^{q_z}(1) \hat{\varphi}_{z}^{p_z}(1) |\mathcal{F}_{i,z}|
\end{equation}
due to the basis function scaling~(\ref{eq:orthogonalBasis}).
The second integral term can be expressed as
\begin{equation}
  \label{eq:faceInterfaceIrregular}
  \int\limits_{\mathcal{I}_{ik|j}}\varphi_i^{q_i}\varphi^{p_k}_k \, \dd^2 {\bf r} = 
  \hat{\varphi}_z^{q_z}(1)\hat{\varphi}_z^{p_z}(-1)
  \int\limits_{x_k \cap x_i} \varphi_{x,i}^{q_x} \,\varphi_{x,k}^{p_x}\, \dd x
  \int\limits_{y_k \cap y_i} \varphi_{y,i}^{q_y} \,\varphi_{y,k}^{p_y}\, \dd y.
\end{equation}
In this case the orthogonality property of the basis functions is lost due to non-identical supports
of $\varphi_i$ and $\varphi_k$.
We gather the terms (\ref{eq:faceInterfaceRegular}) and
(\ref{eq:faceInterfaceIrregular}) in the interior and exterior flux matrices $\mathbf{F^-}$ 
and $\mathbf{F^+}$. Following to the usual notation the sign indicates the
evaluation from the interior and exterior side of the interface.
Any non-regularity of the grid is now concealed within $\mathbf{F}^+$, which reduces to the standard
form on regular grids.

\begin{figure}[htbp]
  \centering
  \includegraphics[width=0.6\textwidth]{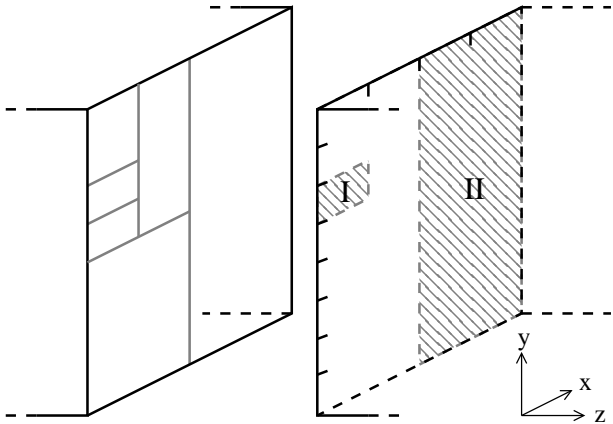}
  \caption{Example of a non-matching interface. Black lines indicate edges of the root tesselation, gray lines indicate
  edges of refined elements. In the figure the interfaces are separated along the $z$-axis for a better visualization.
  The left hand root element has been refined several times, the right hand element is at root level. The interface I connects
  an element of refinement levels $(2, 3, L_z)$ with the root level element. The tick marks indicate possible locations for
  the imprint of elements of these refinement levels. The actual imprint on the root element face
  fills the first and sixth slab along the $x$- and $y$-axis, respectively. The interface II fills the respective second slab
  along the $x$-axis.}
  \label{fig:nonregularInterface}
\end{figure}

For high level hanging nodes the number of integrals to compute quickly becomes large, imposing
a heavy computational burden if integration is performed at run time. However, as the integrals
$\int_{d_k \cap d_i} \varphi_{d,i}^{q_d} \,\varphi_{d,k}^{p_d}\, \dd r_d$
in (\ref{eq:faceInterfaceIrregular}) do not include the actual approximation but basis functions only, they can be 
precomputed analytically (making use of the master basis functions)
and stored in tabulated form in the code. This has to be done for all combinations
of $p_d$ and $q_d$ as well as for each possible edge overlap according to the respective difference
in the refinement levels $\Delta L_d$ (cf.~Fig.~\ref{fig:nonregularInterface}). The number of possible overlaps grows as $2^{\Delta L_d}$.
We tabulated the integrals up to $\Delta L_d = 6$ and for basis functions up to order six, yielding
247 matrices $\mathbf{I}_{\Delta L}$ of size $7 \times 7$. In the isotropic refinement case $\Delta L_d = 6$ corresponds to
one element interfacing with $(2^6)^2 = 4096$ neighbors. In the case of even larger differences in the refinement
levels of neighboring elements, which are unlikely to occur a numerical integration
is invoked at run time. If the neighboring element has a smaller instead of higher refinement level the respective transposed
matrix $(\mathbf{I}_{\Delta L})^\text{T}$ is applied. For upwind fluxes, the interior and exterior flux
matrices do not change, however, they are applied to both, the electric and the magnetic field due to~(\ref{eq:upw}).

In order to further enhance computational performance, all combinations of
$\hat{\varphi}^{q_d}(\pm 1)\hat{\varphi}^{p_d}(\pm 1)$ and the integrals
$\int_{\hat{\mathcal{T}}_d} (\frac{\dd}{\dd \hat{r}_d}  \hat{\varphi}^{q_d} )\hat{\varphi}^{p_d} \,\dd \hat{r}_d$
arising form the stiffness terms of (\ref{eq:DGMaxwell})
are evaluated and tabulated as well. 
Precomputing the interface integrals maintains the high computational efficiency of the
DG methods also for non-regular grids. Using matrix notation, the semi-discrete DG Maxwell
equations~(\ref{eq:DGMaxwell}) read
\begin{equation}
  \label{eq:maxwellSemiMatrix}
  \frac{\dd}{\dd t}
  \begin{pmatrix}
    {\bf M}_\mu \hbold \\
    {\bf M}_\epsilon \ebold
  \end{pmatrix}
  =
  \left(
    \begin{array}{cc}
      \gamma ( {\bf F}^- + {\bf F}^+ )\, {\bf Z} & - {\bf S} + ( {\bf F}^- + {\bf F}^+ ) \\
      {\bf S} - ( {\bf F}^- + {\bf F}^+ ) & \gamma ( {\bf F}^- + {\bf F}^+ ) / {\bf Z}
    \end{array}
  \right)
  \begin{pmatrix}
    \hbold \\
    \ebold
  \end{pmatrix},
\end{equation}
where \textbf{S} and \textbf{Z} denotes the stiffness and impedance
matrix. The matrix operator on the right hand side of~(\ref{eq:maxwellSemiMatrix})
represents a weak DG curl operator. Choosing $\gamma$ as either zero or one yields
centered or upwind fluxes, respectively. By applying centered fluxes the Hamiltonian structure of
Maxwell's equations in continuum is preserved, whereas upwind fluxes lead to a mixed form.
Symplectic explicit time integration can be applied in the former case but not in the latter one~\cite{Schnepp:2010JCP}.
For examples of symplectic time integration for Maxwell's equations in the DG framework see, e.g.,
\cite{Fezoui2005,Gjonaj2006,Canouet2005}.
In \cite{Cockburn2001,Hesthaven2002Maxwell} upwind fluxes and Runge-Kutta schemes are applied for the time integration,
where the latter one is concerned with Maxwell's equations.

\section{Local refinement techniques}
\label{sec:local-refin-techn}

The adaptation techniques presented in the following are based on projections
between the finite element spaces introduced in (\ref{eq:DGtensorSpace}).
The projection operators have been introduced in~\cite{Schnepp:2011RS},
however, they are included for completeness. Also, we address the issue
of stability in depth and amended this section
with examples.

The approximation $f_h$ to a given function $f$ in the FES $\mathcal{V}^P$ is obtained by
performing an orthogonal projection. The projection is carried out in an element-wise manner, by means
of the projection operator $\Pi^p$ given by
\begin{equation}
  \label{eq:projectionOperator}
  f_i = \sum_p \Pi^p (f)_{\mathcal{T}_i} \,  \varphi_i^p =
  \sum_p\frac{\left(\varphi_i^p, f \right)_{\mathcal{T}_i} }{\left(\varphi_i^p, \varphi_i^p \right)_{\mathcal{T}_i}}
  \varphi_i^p,
\end{equation}
where $(u,v)_{\mathcal{T}_i}$ denotes the inner product $\int_{\mathcal{T}_i} uv\, \dd {\bf r}$ on the element $\mathcal{T}_i$ with
the associated 2-norm $(u,u)_{\mathcal{T}_i} = \|u\|_{\mathcal{T}_i}^2$.
When applied successively to all elements and all components of given initial conditions of the electric field,
${\bf E}(t=t_0)$, and the magnetic field, ${\bf H}(t=t_0)$, the respective
DG approximations $\Ebold_h$ and $\Hbold_h$ are obtained.
These approximations are optimal in the sense that
the projection errors $\mathcal{E}_d = U_d - U_{d,h}$
are orthogonal to the space of basis functions $\mathcal{V}^P$
\begin{equation}
  \label{eq:DGapproxError}
  \left( \mathcal{E}_d, \varphi_i^p \right)_{\mathcal{T}_i} = 0; \quad \forall p \in [0,P],\, \varphi_i^p = \hat{\varphi}^p \circ G_i^{-1},\, \varphi^p \in \mathcal{V}^P
\end{equation}

\subsection{$h$-Refinement}
\label{sec:h-refinement}

As stated above $h$-refinement is achieved by means of element bisections along the
coordinate directions, where we allow for anisotropic refinements. The refined
elements are referred to as the left and right hand side element $\mathcal{T}\ileft$ and $\mathcal{T}\iright$
with basis functions denoted as $\hat{\varphi}^{l}_\text{l}$ and $\hat{\varphi}^{r}_\text{r}$
spanning the spaces $\mathcal{V}_{\text{l}}^{L}$ and $\mathcal{V}_{\text{r}}^{R}$ in a full analogy to $\mathcal{V}^P$
defined in~(\ref{eq:DGtensorSpace}).
The approximation orders $L_d$ and $R_d$ in each child element
do not have to be identical, neither are they required to be equal to the respective order $P_d$
of the parent element. The direct sum of the spaces $\mathcal{V_\text{l}}$ and $\mathcal{V_\text{r}}$
is denoted by $\mathcal{V}^+$
\begin{equation}
  \label{eq:legendreSpaceUnion}
  \mathcal{V}^+ = \mathcal{V_\text{l}} \oplus \mathcal{V_\text{r}}.
\end{equation}

In the following, the projection~\eqref{eq:projectionOperator} can be applied
in order to project an approximation given in an element $\mathcal{T}_i$ to the FES associated
with an $h$-refined or $h$-reduced element.
For $h$-refinement this yields
\begin{equation}
  \label{eq:projSolutionRefine}
  ({\bf u}_i)^l_\text{l} = 
  \Pi^l\ileft ({\bf U}_i)_{\mathcal{T}\ileft}, \quad
  ({\bf u}_i)^r_\text{r} =
  \Pi^r\iright ({\bf U}_i)_{\mathcal{T}\iright}.
\end{equation}
Due to the tensor product character of the basis, this can be expressed as
\begin{equation}
  \label{eq:projSolutionRefine2}
  ({\bf u}_i)^l_\text{l} = \sum_p {\bf u}_i^p \,
  \Pi^l\ileft (\varphi_i^p)_{\mathcal{T}\ileft}
  = \sum_p {\bf u}_i^p\,
  \Pi^{l_x}\ileft (\varphi_i^{p_x})_{\mathcal{T}_{x,\text{l}}} \,
  \Pi^{l_y}\ileft (\varphi_i^{p_y})_{\mathcal{T}_{y,\text{l}}} \,
  \Pi^{l_z}\ileft (\varphi_i^{p_z})_{\mathcal{T}_{z,\text{l}}}  
\end{equation}
for the left and right child, respectively. If refinement is carried out along one coordinate only, e.g.~$x$,
this further simplifies to
\begin{equation}
  \label{eq:projSolutionRefine3}
  ({\bf u}_i)^l_\text{l} = 
  \delta_{l_y p_y} \delta_{l_z p_z} \sum_{p_x} {\bf u}_i^p \,
  \Pi^{l_x}\ileft (\varphi_i^{p_x})_{\mathcal{T}_{x,\text{l}}} = 
  \delta_{l_y p_y} \delta_{l_z p_z} \sum_{p_x} {\bf u}_i^p \,
  \frac{\big(\varphi_\text{l}^{l_x}, \varphi_i^{p_x} \big)_{\mathcal{T}_{x,\text{l}}}}{|\mathcal{T}_{x,\text{l}}|},
\end{equation}
where $\delta$ denotes the Kronecker delta. Note that above we loop over $p_x$, whereas in
(\ref{eq:projSolutionRefine2}) the loop parameter is $p$.
As, moreover, $(\varphi_\text{l}^{l_x}, \varphi_i^{p_x})_{\mathcal{T}_{x,\text{l}}}$ vanishes
for any $p_x < l_x$, we can limit the above loop to the range
$[l_x, P_{x,i}]$, which reduces the number of addends to the minimum possible.

For the merging of elements,
the approximation within the parent element, $\mathcal{T}_i$, is considered to be given piece-wise
within its child elements. The projection reads
\begin{equation}
  \label{eq:projSolutionReduce}
    {\bf u}_i^p =
    \Pi^p \big( ({\bf U}_i)_\text{l} + ({\bf U}_i)_\text{r} \big)_{\mathcal{T}_i} =
    \Pi^p \big( ({\bf U}_i)_\text{l} \big)_{\mathcal{T}_i} + \Pi^p \big(({\bf U}_i)_\text{r} \big)_{\mathcal{T}_i},
\end{equation}
where the simplifications~(\ref{eq:projSolutionRefine2}) and (\ref{eq:projSolutionRefine3}) apply.



\subsection{$p$-Refinement}
\label{sec:p-refinement}

For the case of $p$-enrichments, the local FES are
amended with the $(P_d+1)$ order basis functions
\begin{equation}
  \label{eq:pEnrichment}
  \mathcal{V}_i^{P+1} =  \mathcal{V}_i^P \cup \{\varphi_{d,i}^{P_d+1}\},
\end{equation}
 where any (non-zero) number of the local maximum approximation orders $P_d$ may be increased.
Also, an enrichment by more than one higher order basis function is possible.
Formally, we perform the orthogonal projection~(\ref{eq:projectionOperator}),
however, due to the orthogonality property of the basis functions
the coefficients ${\bf u}_i^{0..P}$ remain unaltered under a projection
from $\mathcal{V}_i^{P}$ to $\mathcal{V}_i^{P+1}$. 
Practically, we simply extend the local vectors of coefficients $\mathbf{u}_i$ with the
new coefficients ${\bf u}_i^{P+1}$, which are initialized to zero.

Conversely, for the case of a $p$-reduction, the local FES is reduced by discarding the $P_d$-order basis functions
\begin{equation}
  \label{eq:pReduction}
  \mathcal{V}_i^{P-1} = \mathcal{V}_i^P \thickspace \backslash \thickspace \{\varphi_{d,i}^{P_d}\}.
\end{equation}
Again, by virtue of the orthogonality, we find that the coefficients ${\bf u}_i^{P}$
are deleted from the local vectors of coefficients while the coefficients ${\bf u}_i^{0..P-1}$
remain unaltered.

We denote by $\Pi_\mathcal{T}$ the projection of the global approximation $(\Ebold_h, \Hbold_h)$
from the current discretization to another one obtained by local $h$- and $p$-adaptations.

\subsection{Optimality, Efficiency and Stability}
\label{sec:stability}

\subsubsection{Optimality}
\label{sec:optimality}

An approximation ${\bf U}_h$ with coefficients according to
\eqref{eq:projectionOperator} is optimal in the sense of~(\ref{eq:DGapproxError}).
The approximations within refined and merged elements with coefficients obtained through
the orthogonal projections
\eqref{eq:projSolutionRefine} and \eqref{eq:projSolutionReduce} are, hence, optimal
in the same sense. 

\begin{figure}[htb]
    \centering
    \psfrag{A}[cc]{(a)}
    \psfrag{B}[cc]{(b)}
    \psfrag{a}[cc]{\footnotesize $-1.0$}    
    \psfrag{c}[cc]{\footnotesize $-0.5$}    
    \psfrag{e}[cc]{\footnotesize $0.0$}    
    \psfrag{n}[cc]{\footnotesize $0.5$}    
    \psfrag{o}[cc]{\footnotesize $1.0$}    
    \psfrag{x}[lb]{$x$}
    \psfrag{y}[cb]{\large $\frac{U(x)}{\text{a.u.}}$}
    \includegraphics[width=0.95\textwidth]{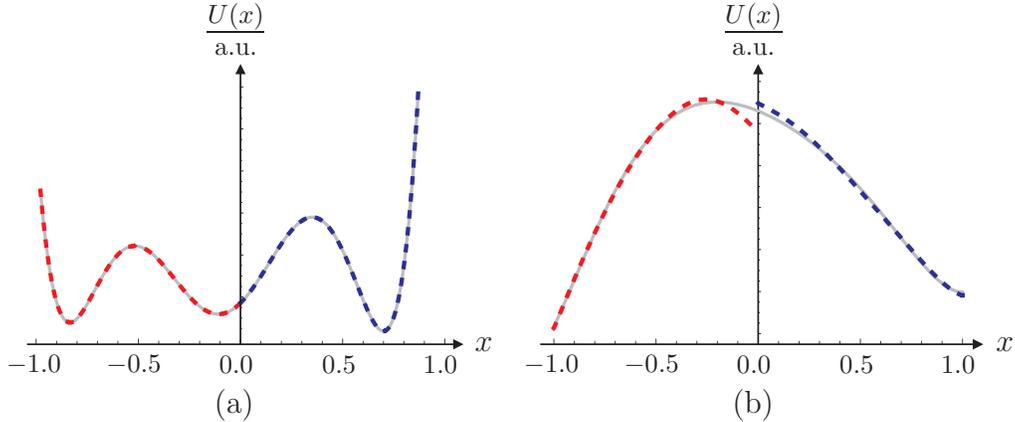}
  \caption{
    Projection based grid refinement and coarsening in one dimension.
    In (a) the projection of a given approximation (gray) to the left and right hand side child elements and
    the respective obtained approximations are shown (dashed, red/blue).
    If $L_x \ge P_x$ and $R_x \ge P_x$, the approximations of the parent and child elements
    agree point-wise. The projection to a merged element shown
    in (b) can, in general, not be exact due to the discontinuity.
  }
  \label{fig:AdapLegendreRemarks}
\end{figure}

If $L_d \ge P_d$ and $R_d \ge P_d$ holds true for all $d$,
the FES $\mathcal{V}$ is a subspace of $\mathcal{V}^+$ (cf.~(\ref{eq:legendreSpaceUnion})) and every
function of $\mathcal{V}$ is representable in $\mathcal{V}^+$ but not
vice versa.
In this case, a given approximation is exactly represented within an element under
$h$-refinement but not under $h$-reduction. See Fig.~\ref{fig:AdapLegendreRemarks}
for an example.

\subsubsection{Efficiency}

Since the projections~(\ref{eq:projSolutionRefine}) for performing $h$-refinement
are independent of the actual approximation,
we also tabulated the projection operators $\Pi\ileft^{l_d}$ and $\Pi\iright^{r_d}$
(expressed in master basis functions)
yielding the matrix operators $({\bf \Pi}\ileft)^+$ and $({\bf \Pi}\iright)^+$,
where the superscript denotes that the refinement level $L$ is increased.
Accordingly, we make use of the matrix operators $({\bf \Pi}\ileft)^-$ and $({\bf \Pi}\iright)^-$
for evaluating the projections of~(\ref{eq:projSolutionReduce}) in the case of element merging.
The matrix operators are related as
\begin{equation}
  \label{eq:projMatricesRelation}
  ({\bf \Pi}\ileft)^- = 2({\bf \Pi}\ileft^+)^\text{T}, \quad ({\bf \Pi}\iright)^- = 2({\bf \Pi}\iright)^\text{T}.
\end{equation}
This allows for the computation of the approximations
within adapted elements by means of efficient matrix-vector multiplications.
As all projection matrices are triangular the evaluation can be carried out as an
in-place operation requiring no allocation of temporary memory.

\subsubsection{Stability}

The global approximation associated with an adapted grid is computed as
$(\Pi_\mathcal{T} \Ebold_h, \Pi_\mathcal{T} \Hbold_h)$. It can be considered
as initial conditions applied on the new discretization obtained by performing the refinement operations.
Assuming stability of the time stepping scheme (cf.~\cite{Hesthaven2002Maxwell,Fezoui2005,Gjonaj2006}),
it is sufficient to show that the application of the projection operators 
at some time $t^*$ does not increase
the electromagnetic energy associated with the approximate DG solution, i.e.,
\begin{equation}
  \label{eq:stabilityCond}
  W_h (\Ebold_h(t^*), \Hbold_h(t^*)) \ge W_h (\Pi_\mathcal{T} \Ebold_h(t^*), \Pi_\mathcal{T} \Hbold_h(t^*)).
\end{equation}
In this case it follows $W_h(t_0)) \ge W_h (t^*)\ge W_h (T)$ and, thus, stability of the adaptive scheme.

Following~(\ref{eq:energyDensity}) the energy associated with element $\mathcal{T}_i$ is given as
\begin{equation}
  \label{eq:DGenergy}
  W_i = \int_{\mathcal{T}_i} \frac{1}{2} \left( \epsilon {\bf {E}}_i^2 + \mu {\bf H}_i^2 \right) \dd^3 {\bf r}
= \frac{1}{2} |\mathcal{T}_i| \left( \epsilon_i \|\mathbf{e}_i\|_2^2 +
      \mu_i \|\mathbf{h}_i\|_2^2 \right).
\end{equation}
As a consequence of (\ref{eq:projSolutionRefine2}),
it is sufficient to show that the energy~(\ref{eq:DGenergy}) is non-increasing
during any adaptation involving one coordinate only.

\paragraph{$h$-Refinement}
\label{sec:h-refinement-1}

For the following discussion of stability it is assumed that
refinement is carried out along the $x$-coordinate. Also we assume
the maximum approximation orders $L,R$ and $P$ to be identical.
It is clarified later, that this does not pose a restriction
to the general validity of the results.

In the case of $h$-refinement, the operators $\mathbf{\Pi}\ileft^+$ and
$\mathbf{\Pi}\iright^+$ project from the space $\mathcal{V}$ to the larger space $\mathcal{V}^+$.
Following the argument of paragraph~\ref{sec:optimality} on optimality, any function defined in
the space $\mathcal{V}$ is exactly represented in $\mathcal{V}^+$.
The conservation of the discrete energy is a direct consequence
as the approximation in the parent and child elements are point-wise
identical\footnote{Identical material properties are assumed for the parent and child elements.}.

We find the following relation for the 2-norms of the respective local vectors of coefficients
\begin{equation}
  \label{eq:DGenergy1Drefine}
  \begin{array}{rcl}
    W_i &=& (W_i)\ileft + (W_i)\iright \\
      \frac{\Delta x}{2} \left(
        \epsilon_i \|\mathbf{e}_i\|_2^2 + \mu_i \|\mathbf{h}_i\|_2^2 \right)
      &=& \frac{\Delta x}{4}  \left(
        \epsilon_i \|(\mathbf{e}_i)\ileft\|_2^2 +
        \mu_i \|(\mathbf{h}_i)\ileft\|_2^2
        + \epsilon_i \|(\mathbf{e}_i)\iright\|_2^2 +
        \mu_i \|(\mathbf{h}_i)\iright\|_2^2 \right) \\
       2 \left(
        \epsilon_i \|\mathbf{e}_i\|_2^2 + \mu_i \|\mathbf{h}_i\|_2^2 \right)
      &=& \epsilon_i (\|(\mathbf{e}_i)\ileft\|_2^2 + \|(\mathbf{e}_i)\iright\|_2^2 )
         + \mu_i ( \|(\mathbf{h}_i)\ileft\|_2^2 + \|(\mathbf{h}_i)\iright\|_2^2 ).
  \end{array}
\end{equation}
The exemplary parent element approximation plotted in Fig.~\ref{fig:AdapLegendreRemarks}a has a maximum
order of $P = 6$ with all coefficients equal to one. The coefficients of the 
child element approximations and the square values of their 2-norms are given in Tab.~\ref{tab:AdapDGcoeffsR}.
If the vector $\mathbf{u}$ is considered to be either the vector of coefficients of the electric
field $\mathbf{e}$ or the magnetic field $\mathbf{h}$
the result agrees with~(\ref{eq:DGenergy1Drefine}).

\begin{table}[htbp]
  \centering
  \begin{tabular}[H]{c||ccccccc|r}
   & $u_0$ &  $u_1$ &  $u_2$ &  $u_3$ & $u_4$ &  $u_5$ & $u_6$ & $\| \mathbf{u} \|_2^2$ \\
    \hline\hline
   $U(x)$ & \whitem 1.0000 & \whitem  1.0000 & \whitem  1.0000 & \whitem  1.0000 & \whitem 1.0000 & \whitem  1.0000 & \whitem 1.0000 & 7.0000\\
   $U(x)\ileft$ & \whitem 0.2574 &         -0.1355 &         -0.2606 &         -0.1446 & \whitem 0.4276 &         -0.1563 & \whitem 0.0156 & 0.3808\\
   $U(x)\iright$ & \whitem 1.7426 & \whitem  1.5631 & \whitem  1.6819 & \whitem  2.0399 & \whitem 1.0494 & \whitem  0.2181 & \whitem 0.0156 & 13.6192
  \end{tabular}
  \caption{Parent and child element coefficients of the function plotted in Fig.~\ref{fig:AdapLegendreRemarks}a}
  \label{tab:AdapDGcoeffsR}
\end{table}

The $h$-coarsening operators $\mathbf{\Pi}\ileft^-$ and $\mathbf{\Pi}\iright^-$ project a function from the space
$\mathcal{V}^+$ to the smaller space $\mathcal{V}$. Since $\mathcal{V}$ is a subspace
of $\mathcal{V}^+$, it is immediately concluded that, in general, energy is lost during
the coarsening process. The discrete energy can only be preserved if the union of the left and right hand functions is
an element of the smaller space $\mathcal{V}$. Starting with the coefficients of the child elements,
given in Tab.~\ref{tab:AdapDGcoeffsR}, the parent element coefficients are exactly recovered and
the discrete energy is preserved.

In particular, it can be shown from algebraic properties of the projection matrices, that the
discrete energy for arbitrary fine grid coefficients
is always non-increasing during $h$-coarsening. First, the $(P_d \times 2P_d)$
projection matrix $\mathbf{\Pi}^-$ is defined as
\begin{equation}
  \label{eq:AdapDGprojectionCoarsenTotal}
  \mathbf{\Pi}^- = \Big( \mathbf{\Pi}^-\ileft \,\, \mathbf{\Pi}^-\iright \Big).
\end{equation}
The coefficients of the child elements are gathered in one vector $(\mathbf{u}_i)^+$
\begin{equation}
  \label{eq:AdapDGcoeffsTotal}
  (\mathbf{u}_i)^+ = 
  \begin{pmatrix}
    (\mathbf{u}_i)\ileft\\ (\mathbf{u}_i)\iright
  \end{pmatrix}.
\end{equation}
Then, the coefficients of the parent element are given as
\begin{equation}
  \label{eq:AdapDGcoeffsCoarseTotal}
  \mathbf{u}_i = \mathbf{\Pi}^- (\mathbf{u}_i)^+,
\end{equation}
which is equivalent to Eqn.~(\ref{eq:projSolutionReduce}).
Using~(\ref{eq:DGenergy}) and (\ref{eq:DGenergy1Drefine}), the following must hold true in order
to guarantee a non-increasing discrete energy
\begin{equation}
  \label{eq:AdapDGenergyHCoarseningCond}
  \begin{array}{rcl}
    2 \| \mathbf{u}_i \|_2^2  & \stackrel! \le&  \| (\mathbf{u}_i)^+ \|_2^2 \\
    2 \, \mathbf{u}_i^\text{T} \mathbf{u}_i  & \stackrel! \le&  \left((\mathbf{u}_i)^+\right)\transp (\mathbf{u}_i)^+ \\
    2 \, \left(\mathbf{\Pi}^- (\mathbf{u}_i)^+ \right) \transp \mathbf{\Pi}^- (\mathbf{u}_i)^+  & \stackrel! \le&  \left((\mathbf{u}_i)^+\right)\transp (\mathbf{u}_i)^+ \\
    \frac{ \left((\mathbf{u}_i)^+\right)\transp \left(\mathbf{\Pi}^-\right)\transp \mathbf{\Pi}^- (\mathbf{u}_i)^+}
    {\left((\mathbf{u}_i)^+\right)\transp (\mathbf{u}_i)^+} & \stackrel! \le& \frac{1}{2}.
  \end{array}
\end{equation}
In order to fulfill this it is sufficient to demand
\begin{equation}
  \label{eq:AdapDGprojectionCoarseEV}
  \max \left\{ \text{eig}\left(\left(\mathbf{\Pi}^-\right)\transp \mathbf{\Pi}^- \right)\right\} \le \frac{1}{2}.
\end{equation}
Since the matrix  $\big(\left(\mathbf{\Pi}^-\right)\transp \mathbf{\Pi}^-\big)$ has
the $P_d$-times degenerated eigenvalues $1/2$ and $0$ this is always fulfilled.

The coefficients for the example shown in Fig.~\ref{fig:AdapLegendreRemarks}b are listed in
Table~\ref{tab:AdapDGcoeffsC}. The sum of the 2-norms for the left and right hand child vectors of coefficients
yields 1.9003 while twice the value obtained for the parent element evaluates to 1.8996. Thus, energy
was lost during $h$-coarsening.

\begin{table}[htbp]
  \centering
  \begin{tabular}[H]{c||cccc|r}
   & $u_0$ &  $u_1$ &  $u_2$ &  $u_3$ & $\| \mathbf{u} \|_2^2$ \\
    \hline\hline
    $U(x)$        & 0.9500 &        -0.0433 & -0.2073 & \whitem 0.0498 & 0.9498 \\
    $U(x)\ileft$ & 1.0000 & \whitem 0.2000 & -0.1000 &        -0.0100 & 1.0501 \\
    $U(x)\iright$ & 0.9000 &        -0.2000 & -0.0100 & \whitem 0.0100 & 0.8502
  \end{tabular}
  \caption{Parent and child element coefficients of the function plotted in Fig.~\ref{fig:AdapLegendreRemarks}b}
  \label{tab:AdapDGcoeffsC}
\end{table}

\paragraph{$p$-Refinement}
\label{sec:p-refinement-1}

In order to show stability of the $p$-adaptation we again consider the energy stored in an element given
by~(\ref{eq:DGenergy}). In the case of $p$-enrichment,
the local vectors of DoF are extended by the coefficients corresponding to the $(P+1)$-order basis functions.
Since these coefficients are initialized to zero, it holds true
\begin{equation}
  \label{eq:AdapDGpenrichEnergy}
  \| (\mathbf{u})^P \|_2^2 = \|(\mathbf{u})^P \|_2^2 + 0
  = \|(\mathbf{u}_{0..P})^{P+1}\|_2^2 + \|(\mathbf{u}_{P+1})^{P+1}\|_2^2 = \|(\mathbf{u})^{P+1} \|_2^2.
\end{equation}
The discrete energy
is exactly conserved.

In the case of a $p$-reduction, the coefficients assigned to the highest order basis functions
are removed from the vectors of DoF. Consequently, it holds true
\begin{equation}
  \label{eq:AdapDGpreduceEnergy}
  \| (\mathbf{u})^{P} \|_2^2 = \| (\mathbf{u}_{0..P-1})^{P} \|_2^2 + \|(\mathbf{u}_{P})^{P}\|_2^2 \le \| (\mathbf{u}_{0..P-1})^{P} \|_2^2 = \| (\mathbf{u})^{P-1} \|_2^2,
\end{equation}
and the discrete energy is either preserved or otherwise reduced. If the decision for reducing the order is a correct one, the highest order
coefficients are small and the induced energy loss is small.

In the discussion of stability for $h$-adaptations it was assumed that the
maximum approximation orders $L,R$ and $P$ are identical. After showing that
$p$-adaptation does not increase the electromagnetic energy either, it can be concluded that this assumption does not restrict
the validity of the results obtained as
the problem can be reduced to performing $h$- and $p$-adaptations sequentially.

Finally, we will make some remarks on mixed $h$- and $p$-adaptations.
If an $h$-coarsening goes along with a $p$-enrichment, the latter should be performed first.
It amounts to a projection to the superspace $\mathcal{V}_h^{P+1}$. 
Projection errors are limited to the projection onto the space $\mathcal{V}_{2h}^{P+1}$.
If the projections are carried out in reversed order the final space is $\mathcal{V}_{2h}^{P+1}$ as well,
however, the second projection from $\mathcal{V}_{2h}^{P}$ adds zeros to the local
vector of DoF only, resulting in an increased overall projection error.
On the contrary, if an $h$-refinement goes along with a $p$-reduction,
the former should be carried out first for the same reason. However, the two adaptations can
be performed simultaneously by employing projection matrices
of the size $(P_d^\text{new} \times P_d^\text{old})$.

\section{Automatic $hp$-adaptivity: Error and smoothness estimation}
\label{sec:test-cases-one}

In the preceding section we presented efficient techniques for performing
local $h$- and $p$-adaptations, which can be considered as the basic toolbox within the
larger frame of an $hp$-adaptive DG method. In this section we address the
critical issues of locally estimating the approximation error and
solution smoothness, which is required for driving the adaptation process.
Note that as we are interested in estimating the smoothness
of the actual solution, the discontinuous nature of the DG approximation is not
a concern. In fact, we will show how the discontinuities at element boundaries
can be exploited for constructing a smoothness indicator.

In order to perform an adaptation of an $hp$-mesh two steps have to be carried out.
First, the elements requiring adaptation have to be identified. This is achieved by
an element-wise error estimation. If the estimate exceeds some tolerance
the element requires refinement, elements having a small error are eligible for coarsening.
In a second step, the kind of adaptation ($h$ or $p$) suitable for the respective element 
has to be decided upon. Here, we distinguish between the refinement and the coarsening case.

For the coarsening case the best option is obtained by consecutively testing possible
de-refinements from a set of candidates~\cite{Demkowicz:1989p1981,Solin:2010p1978}.
The set contains at least the candidates obtained by $h$-reducing the refinement level to $L_i-1$
and by reducing the polynomial order to $P_i-1$ but larger sets of candidates are possible as well.
The list of candidates can be extended hierarchically. If, e.g., reducing the polynomial order
yields an approximation still fulfilling the accuracy requirements, subsequent candidates can be
generated as long as the accuracy demands are met.

An extension to the refinement case is possible as shown in~\cite{Rachowicz:1989id,Demkowicz:1989p1981,Solin:2010p1978},
however, at the price of computing a globally $h$- and $p$-refined solution.
We pursued a different approach based on a local smoothness indicator
(cf.~Sec.~\ref{sec:smoothn-estim}).

\subsection{Error Estimation}
\label{sec:error-estimation}

Error estimation is addressed in a large number of publications out of which
we refer to the introductions \cite{Ainsworth:2000tw,Cockburn2003} and 
references therein. In the context of this paper, we focus on
\cite{Cockburn2003a}, where a relation between the size of the jumps of some quantity at element boundaries
\begin{equation}
  \label{eq:DGjumps}
  \llbracket U \rrbracket_{ik} = U_{i|\mathcal{I}_{ik}} - U_{k|\mathcal{I}_{ik}}
\end{equation}
and the residual within in the respective element is derived for a model problem.
Following this idea, we obtained a similar result for Maxwell's equations, which we apply as an error
estimate. This is a valuable
tool for the development of an adaptive DG method because the evaluation of the
jumps is a computationally inexpensive operation. Moreover, it can be incorporated with
the computation of the fluxes, which renders the extra costs negligible.
The derivation is outlined in the following.

The starting point is equation~\eqref{eq:DGFaraday}, where we integrate the
second term of the volume integral by parts to obtain
\begin{equation}
  \label{eq:estimate1}
    \int\limits_{\Ti} \psi \, \mu \frac{\partial}{\partial t}{\bf H}_h \dd^3 {\bf r}
      + \int\limits_{\partial \Ti} \psi \, \big(\mathbf{n} \times ({\bf E}_h^* - {\bf E}_h)\big)  \, \dd^2 {\bf r}
    + \int\limits_{\Ti} \psi (\mathbf{\nabla}  \times {\bf E}_h ) \dd^3 {\bf r}
    = 0,
\end{equation}
where we returned to denoting the test function by $\psi$ (cf.~\eqref{eq:weakMaxwell})
for emphasizing the freedom of choice. Taking $\psi = 1$ we obtain
\begin{equation}
  \label{eq:estimate2}
    {\bf R}_i = \int\limits_{\Ti}  {\bf R}_h\,  \dd^3 {\bf r}
      = \int\limits_{\partial \Ti}   \, \mathbf{n} \times ( {\bf E}_h - {\bf E}_h^*)  \, \dd^2 {\bf r},
\end{equation}
with the residual
\begin{equation}
  \label{eq:estimateResidual}
  {\bf R}_h = \mu \frac{\partial}{\partial t}{\bf H}_h + (\mathbf{\nabla}  \times {\bf E}_h ),
\end{equation}
and the element error estimate
\begin{equation}
  \label{eq:estimate3}
  \mathcal{E}_i = \|{\bf R}_i\|_{\mathcal{T}_i}.
\end{equation}
The global error estimate is obtained as $\mathcal{E} = \sum_i \mathcal{E}_i$.
By inserting either of the fluxes~\eqref{eq:cent} or \eqref{eq:upw}, i.e.
centered or upwind, in~\eqref{eq:estimate2} we obtain
\begin{subequations}
  \label{eq:estimateFluxes}
  \begin{align}
  \label{eq:estimateFluxesCT}
      {\bf R}_i^{\text{ct}} &= \frac{1}{2} \, \sum_k \int\limits_{\mathcal{I}_{ik}}   \, \mathbf{n} \times \llbracket {\bf E} \rrbracket_{ik}   \, \dd^2 {\bf r},\\
  \label{eq:estimateFluxesUP}
      {\bf R}_i^{\text{up}} &= \frac{1}{2} \, \sum_k \int\limits_{\mathcal{I}_{ik}}   \, \mathbf{n} \times 
      (\llbracket {\bf E} \rrbracket_{ik}  + Z \llbracket {\bf H} \rrbracket_{ik} )
      \, \dd^2 {\bf r},
  \end{align}
\end{subequations}
where $k$ numbers all interfaces of element $i$. This
establishes the link between the jump sizes and the residual.
Note, that there is a consistency with the observation that in homogeneous regions  
the jumps disappear if the numerical solution is exact.

\begin{figure}[bht]
    \centering
    \psfrag{h}[cb]{\footnotesize {$h$}}
    \psfrag{LJ}[cc]{Gaussian}
    \psfrag{LR}[cc]{Trapezoidal}
    \psfrag{LP=1}[lb]{\scriptsize {$\text{L}^2\, P=1$}}
    \psfrag{LP=2}[lb]{\scriptsize {$\text{L}^2\, P=2$}}
    \psfrag{LP=4}[lb]{\scriptsize {$\text{L}^2\, P=4$}}
    \psfrag{JP=1}[lb]{\scriptsize {$\mathcal{E}\, P=1$}}
    \psfrag{JP=2}[lb]{\scriptsize {$\mathcal{E}\, P=2$}}
    \psfrag{JP=4}[lb]{\scriptsize {$\mathcal{E}\, P=4$}}
    \psfrag{L2Error}[lb]{\footnotesize {error}}
    \includegraphics[width=\textwidth]{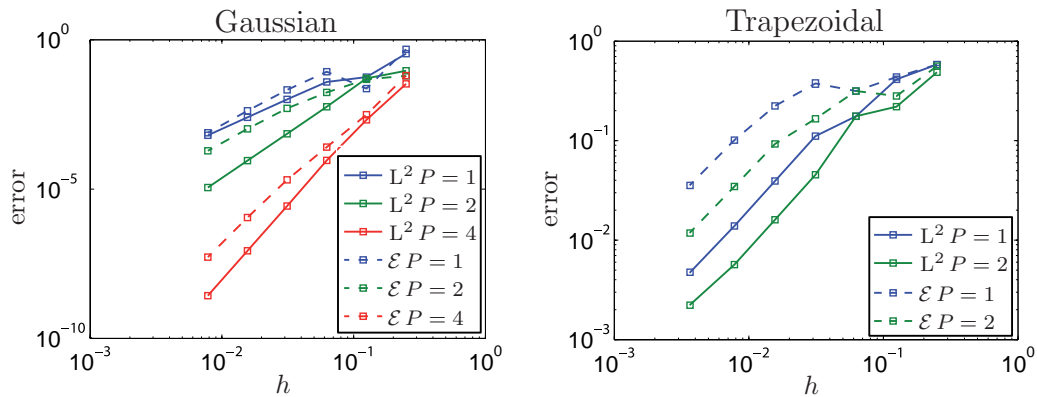}
  \caption{
    Global approximation error for a Gaussian and a trapezoidal wave form under uniform grid refinement.
    Bold lines correspond to the $L^2$-norm, dashed lines indicate the estimate.
  }
  \label{fig:L2vsJump}
\end{figure}

Figure~\ref{fig:L2vsJump} shows results of tests we conducted for
investigating the estimate performance. The tests were carried out
in a one-dimensional domain using a Gaussian and a trapezoidal wave form as examples for
a smooth solution and non-smooth solution. Bold lines in Fig.~\ref{fig:L2vsJump}
correspond to the global approximation error in the $L^2$-norm, dashed lines
represent the global error estimate $\mathcal{E}$.

The estimate performs well for the Gaussian wave form. It tends to overestimate the error, however,
we consider the discrepancies to be acceptable, especially given that it reproduces
the correct trend under grid refinement. The situation is not as good for the trapezoidal
wave form, where the overestimation of the error is more significant. Nevertheless,
also for this example the trend under grid refinement is correct. 

\subsection{Smoothness Estimation}
\label{sec:smoothn-estim}

Once an element is marked for refinement,
it has to be decided upon the kind of adaptation to perform.
This $hp$-decision is based on the solution smoothness.
If the approximation within the element under
consideration is sufficiently smooth we perform $p$-refinement in order to obtain
spectral convergence~\cite{Babuska:1994wj,Schwab:1999tu}. Otherwise, we choose $h$-refinement.
Hence, the discretization error and the achievable order of convergence critically depend on
correct $hp$-decisions. This in turn makes reliable smoothness estimators a necessity for adaptive codes.

The subject of smoothness estimation has a good coverage in the
 literature as well. In one of the first publications on adaptive DG methods \cite{Bey:1996vu}
the authors propose to estimate the local solution regularity based on the decay rate
of the local error. As the local error is unknown and subject to estimation
itself the adaptation procedure critically hinges on the error estimate.
More recent works such as
\cite{Houston:2005p1984,Krivodonova2004,Persson:2006p1988,Casoni:2009p1987,Klockner:2011ia,Wihler:2011jf}
attempt to estimate the solution smoothness based on a variety of 
properties of the local solution, but they do not include the error estimate.
See \cite{Houston:2005p1984} also for a more complete overview of different smoothness estimation
concepts.

A first family of popular methods estimates the local analyticity by
projecting the solution to a set of orthogonal functions (e.g. Legendre polynomials)
and investigates the decay of the coefficients in on way or another. This approach is
being followed by \cite{Houston:2005p1984,Persson:2006p1988,Casoni:2009p1987,Klockner:2011ia}, where the latter one
builds on top of \cite{Persson:2006p1988} and obtains improved results especially when
the sequence of coefficients exhibits a pronounced odd-even behavior. 
As we express the approximation as a Legendre
series within each element the projection step would not be required.
In \cite{Houston:2005p1984,Wihler:2011jf} the authors attempt to estimate
the local Sobolev regularity index directly, where the former one requires
the representation of the solution as a Legendre series as well whereas
the latter one is a novel idea based on continuous Sobolev embeddings
and does not require a series expansion. Finally, smoothness indicators
can also be built from superconvergence properties of the DG method
\cite{Adjerid:2002wb,Krivodonova2004}, which will be described in more
details below.

The first family of smoothness estimates are mainly applied in the context
of shock capturing, specifically for controlling artificial viscosity within
high order DG simulations. In this context they proved to reliably
achieve a good performance. In our experience, however, they
are rather not suited for controlling an $hp$-decision as they require
a minimum number of coefficients for estimating the decay rate,
which might not be available for low order elements.

We adapted the smoothness indicator for hyperbolic conservation laws~\cite{Krivodonova2004}, which
exploits the difference in the convergence rate of interface jumps for smooth and non-smooth solutions.
From our experience this indicator works very robust and, most importantly,
this remains valid down to very low polynomial orders of one and even zero.
This coincides with the experiences reported in \cite{Luo:2009p1985,Richter:2009p1989}.
It should be noted, however, that no estimated value for the local analyticity or regularity
is obtained. We plan to test the estimate \cite{Wihler:2011jf} since it appears
to be working down to low orders while obtaining an estimated regularity index at the same time. 
For the time being, however, we consider an adapted version of the smoothness indicator \cite{Krivodonova2004}.

For a scalar quantity $Q$ that can be a solution component or a derived quantity as well,
it holds true~\cite{Adjerid:2002wb}
\begin{equation}
\label{eq:convout}
\frac{1}{|\mathcal{I}_{ik}|} \int_{\mathcal{I}_{ik}^\text{out}} (Q_i - Q) \dd^2 {\bf r} =\mathcal{O}(h^{2p+1}),
\end{equation}
where $\mathcal{I}_{ik}^\text{out}$ denotes an outflow boundary regarding the quantity $Q$.
Following~\cite{Krivodonova2004} the superconvergence property~\eqref{eq:convout} can be exploited for constructing
a smoothness indicator. To this end we consider
\begin{equation}
  \label{eq:intJmpQ}
  \int_{\mathcal{I}_{ik}^\text{in}} \llbracket Q \rrbracket_{ik} \dd^2 {\bf r} =
  \int_{\mathcal{I}_{ik}^\text{in}} (Q_i - Q_k) \dd^2 {\bf r} =
  \int_{\mathcal{I}_{ik}^\text{in}} (Q_i- Q) \dd^2 {\bf r} +\int_{\mathcal{I}_{ki}^\text{out}} (Q - Q_k ) \dd^2 {\bf r}.
\end{equation}
As a result of~\eqref{eq:convout} the last integral converges as 
$\mathcal{O}(h^{2(p+1)})$ and the left hand side expression is $\mathcal{O}(h^{p+2})$
in regions of smooth solution. If, however, $Q$ is non-smooth in the
vicinity of $\mathcal{I}_{ik}$ then both right hand side integrals are only
$\mathcal{O}(h)$. Normalizing~\eqref{eq:intJmpQ} by an average convergence rate
and the $L^2$-norm of the considered quantity on $\mathcal{T}_i$
yields a smoothness indicator
\begin{equation}
  \label{eq:detector}
  I_{ik} = \frac{|\int_{\mathcal{I}_{ik}^\text{in}} \llbracket Q_i \rrbracket \, \dd^2 {\bf r}|}  {h^{(p+1)/2}|\mathcal{I}_{ik}|\,\|Q_i\|_2 },
\end{equation}
where $h$ is a characteristic measure of the size of element $i$. We use the length of the
edge normal to the interface $\mathcal{I}_{ik}$, which for the one-dimensional case
is simply the element length. For $\mathcal{I}_i > 1$ the solution on $\mathcal{I}_i$ is considered to be non-smooth
and smooth otherwise with $h$-refinement being performed in the former case
and $p$-refinement in the latter one. As the indicator is applied at element faces,
it allows for indicating the solution smoothness along each coordinate separately
which can be expoited for driving anisotropic $hp$-refinement.

We use components of the Poynting vector ${\bf S} ={\bf E} \times {\bf H}$
as the considered quantity $Q$.
Then, inflow boundaries are recognized by ${\bf S} \cdot \mathbf{n}_{ik} < 0$.
The Poynting vector represents the energy flux density.
As a consequence of this particular choice, the nominator of (\ref{eq:detector}) is large
for big jumps of the energy flux density across the element boundary.
This is likely to indicate a low regularity of the local field solution and
allows for interpreting the smoothness indicator based on the underlying
physics. As this interpretation does not depend on the element order,
we assume that this explains our observation of a good performance
also for low orders (see Example~\ref{sec:autom-adapt-one}).

\section{Application examples}
\label{sec:application}

In this section we will present two examples of the fully automatic $hp$-adaptive
solution of wave propagation problems in one-dimensional space and
one example which proves the capability of our implementation to handle
large problems in three-dimensional space including thousands of 
mesh adaptations. However, in the latter example we drive the adaptation using
a simple energy density criterion for the practical reason that 
the implementation of the estimates for the three-dimensional case is not yet completed.

\subsection{Automatic adaptation in one-dimensional space}
\label{sec:autom-adapt-one}

\begin{figure}[htb]
    \centering
    \psfrag{x}[cc]{\normalsize {$x$ / a.u.}}
    \psfrag{EH}[cb]{\footnotesize {$E_y$, $\sqrt{\mu/\epsilon}\, H_z$, $P/10$}}
    \psfrag{G}[ct]{Gaussian}
    \psfrag{T}[ct]{Trapezoidal}
    \psfrag{00}[rc]{\footnotesize 0.0}
    \psfrag{01}[rc]{\footnotesize -1.0}
    \psfrag{10}[rc]{\footnotesize 1.0}
    \psfrag{02}[rc]{\footnotesize -2.0}
    \psfrag{20}[rc]{\footnotesize 2.0}
    \includegraphics[width=\textwidth]{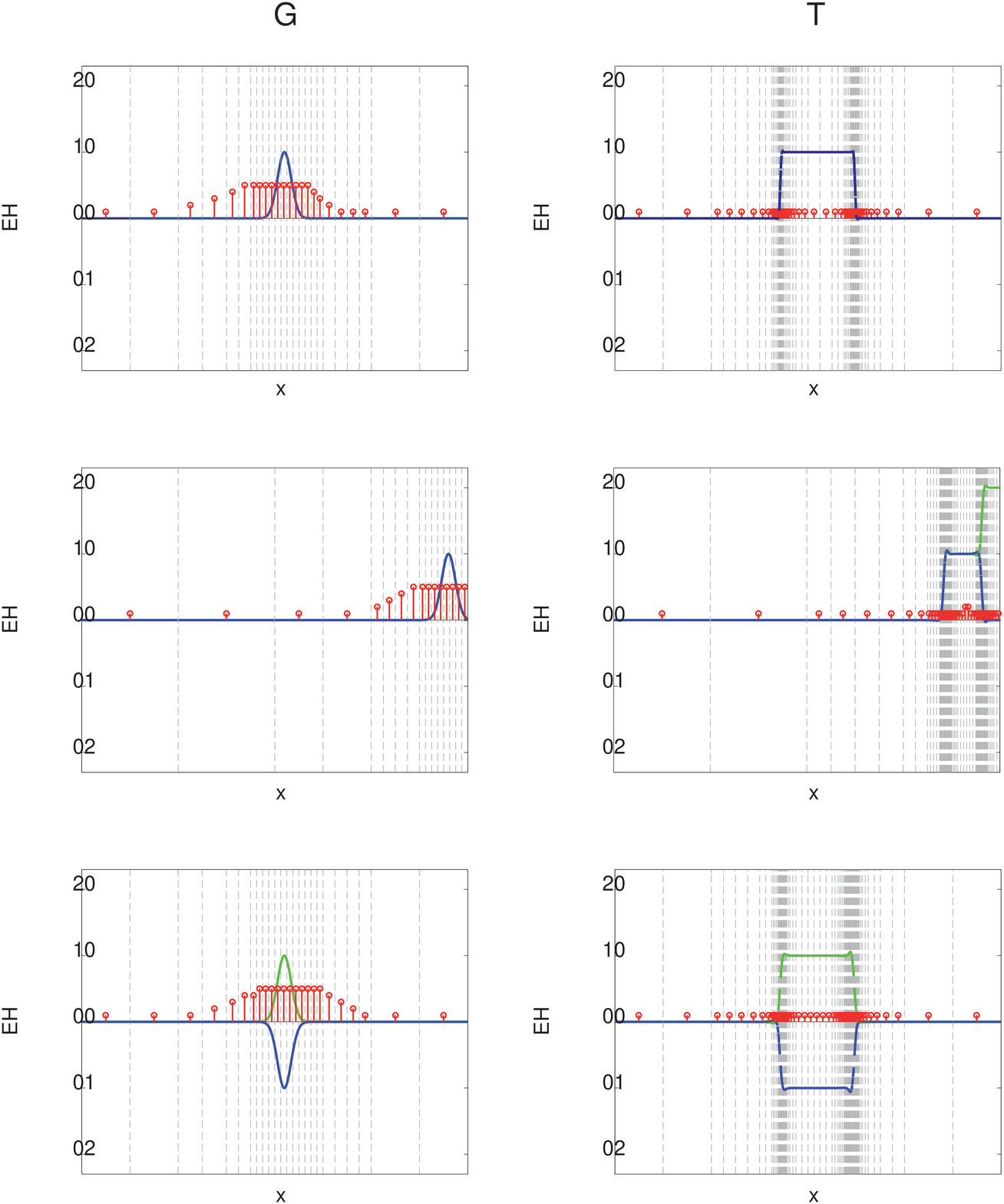}
  \caption{
    Simulation of a {Gaussian} and trapezoidal wave packet using automated $hp$-adaptation.
    The electric and magnetic field are plotted in blue and green, respectively.
    Gray dashed lines indicate grid node positions, red circles indicate the polynomial
    order of the respective element (divided by ten). The polynomial order is bound in between one and five.
    The initial, an intermediate and the final solutions are shown from top to bottom.
    For the {Gaussian} packet, the $h$-refinement level does not exceed two,
    while taking full advantage of the highest order in the packet region.
    For the trapezoidal packet, an order of two is not exceeded.
    However, the algorithm makes use of $h$-refinement levels up to six.
  }
  \label{fig:TestsHPDG}
\end{figure}

The adaptation strategy described above is being tested with a Gaussian and a trapezoidal
wave packet in a one-dimensional setting as shown in Fig.~\ref{fig:TestsHPDG}.
The error tolerances are $10^{-3}$ for the former and $10^{-1}$ for
the latter case. The gray dashed lines depict the position of the grid points and the
red circles indicate the approximation order employed
for the respective element divided by ten. For the Gaussian packet
the adaptation algorithm chooses medium sized to big elements and the
preset maximum approximation order of five in the vicinity of the packet.
For the non-smooth trapezoidal packet, the maximum approximation 
order chosen by the algorithm throughout the simulation is two.
In the vicinity of the pulse edges a high degree of $h$-refinement
is applied, thus showing the desired behavior for the second packet as well.

\begin{figure}[htb]
    \centering
    \includegraphics[width=\textwidth]{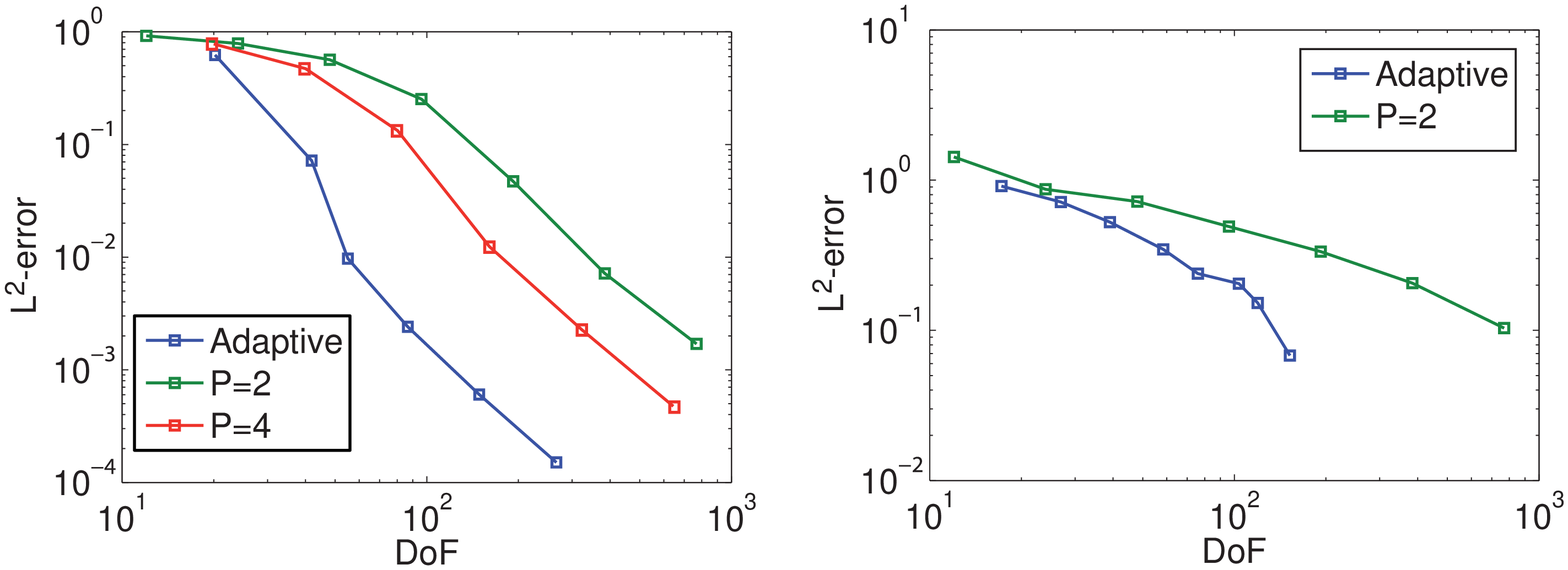}
    \caption{$L^2$-error of the solution vs.~number of DoF for Gaussian (left) and trapezoidal wave packet (right)
      as depicted in Fig.~\ref{fig:TestsHPDG}. For the adaptive simulations the number of DoF corresponds to the
      mean value of all time steps.}
  \label{fig:TestsHPDG_error}
\end{figure}

Figure~\ref{fig:TestsHPDG_error} shows a comparision of the $L^2$-error achieved
for varying numbers of DoF using $hp$-adaptivity (corresponding to different
error tolerances) and fixed meshes of uniform polynomial order. Using $hp$-meshes
the number of DoF required for obtaining a certain accuracy is clearly reduced.
The achievable gain by using $hp$-adaptivity, however, strongly depends on
the problem under consideration. For the examples considered here, e.g., enlarging the domain
while preserving the size of the wave packet will increase the separation between the
curves corresponding to the adaptive and non-adaptive solutions in favor of
the $hp$-solution
and vice versa. In other words, the more pronounced the multiscale character
of the problem is the higher is the gain by applying $hp$-adaptivity.

\subsection{Proof of feasibility example in three-dimensional space}
\label{sec:proof-feas-example}

In this section we consider the backscattering of a wide-band electromagnetic pulse from a passive radar reflector.
As stated above, the implementation of the error and smoothness estimates for applications 
in three-dimensional space is underway, and we resorted to applying a physics based 
criterion for driving the mesh adaptation process.
The purpose of this example is to demonstrate the ability of our implementation~\cite{SMOVE}
to handle large scale problems using $hp$-meshes. This involves the handling of meshes
with hanging nodes and thousands of adaptations on the element level, which are carried out by
means of the efficient techniques presented above. 

\begin{figure}[tbhp]
  \begin{boxedminipage}[h]{\textwidth}
  \centering
  \vspace{5mm}
  \includegraphics[width=0.9\textwidth]{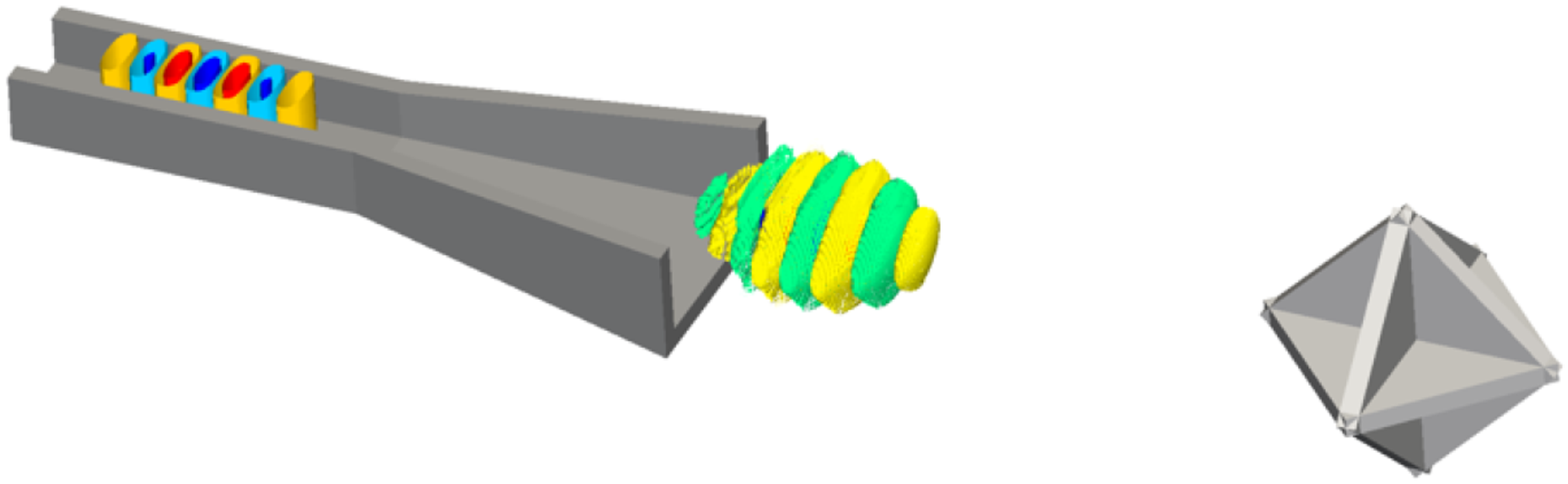}
  \vspace{5mm}
  \end{boxedminipage}
  \caption{Scattering setup consisting of a horn antenna (in cut view) illuminating a radar reflector. The antenna emits
    a broadband electromagnetic waveform, whose electric field contours are depicted at two instances in time.}
  \label{fig:setup}
\end{figure}

In this example
a radar reflector is illuminated off-center by a horn antenna, which emits a Gaussian-modulated sinusoidal pulse
covering the frequency range from 20 to 30 GHz. The initial waveform is a TE$_{10}$ (transverse electric) wave.
The setup is depicted in Fig.~\ref{fig:setup}, where the antenna is shown in a cut view along with contour plots
of the pulse at two instances in time ($t = 1$ ns and $t = 10$ ns). Table~\ref{tab:parameters} lists
the parameters and dimensions. We simulated the full scattering
process starting from the excitation inside the waveguide to the recording of the reflected
fields at the same position. The total propagation distance is about sixty wavelengths.

\begin{table}[bth]
  \centering
  \begin{tabular}[h]{ll}
    \hline\hline
    Parameter                  & Value \\
    \hline
    Waveform mode & TE$_{10}$\\
    Waveform frequency range & 20 -- 30 GHz\\
    Waveguide type & WR-62 \\
    Waveguide width   & 15.8 mm\\
    Waveguide height  & 7.9 mm\\
    Horn width    & 39.7 mm \\
    Horn height   & 29.0 mm \\
    Horn depth    & 67.5 mm \\
    \hline\hline
  \end{tabular}
  \caption{Setup parameters and dimensions}
  \label{tab:parameters}
\end{table}

We chose a maximum $h$-refinement level of two and the local
element order to vary in between zero and four. The local energy density, introduced
in~(\ref{eq:energyDensity}), serves as the criterion for controlling the adaptation procedure.
Denoting by $\widebar{w}_i$ the average energy density of element $i$ 
and by $w_i = \widebar{w}_i/\hat{w}$ the normalized energy density with
$\hat{w} = \underset{i}{\max}\, \left\{\widebar{w}_i\right\}$, we assigned the local refinement levels according to
$w_i < 0.5\delta: L_{i} = 0;  w_i \in [0.5\delta,\delta): L_i = 1 ; \delta \le w_i: L_i =2$
and polynomial orders as
$w_i < 0.5 \delta: P_{i} = 0;  w_i \in [0.5\delta, \delta): P_i = 1 ; w_i \in [\delta,2\delta): P_i =2;  w_i \in [2\delta,3\delta): P_i = 3 ;  w_i \in [3\delta,4\delta): P_i =4$ with $\delta = 0.01$.

The initial discretization consisted of $45 \times 35 \times 100 = 157,500$
elements. During the simulation the number of elements varies and grew
strongly after scattering from the reflector took place, when it
reached close to 800,000 elements corresponding to
slightly more than 55~million DoF.
For comparison, we note that employing the finest mesh resolution globally
as well as fourth order approximations uniformly would lead to approximately 7.5~billion ($10^9$) DoF.
This corresponds to a factor of approximately 130 in terms of memory savings.
We emphasize that the simulations were carried out on a single machine.
The implementation takes full advantage of multi-core capabilities through OpenMP parallelization.
The numerous run-time memory allocations and deallocations are handled through a
specialized memory management library based on memory blocking,
which we implemented for supporting the main code~\cite{DynaMO}.

\begin{figure}[tbhp]
  \centering
  \includegraphics[width=1.0\textwidth]{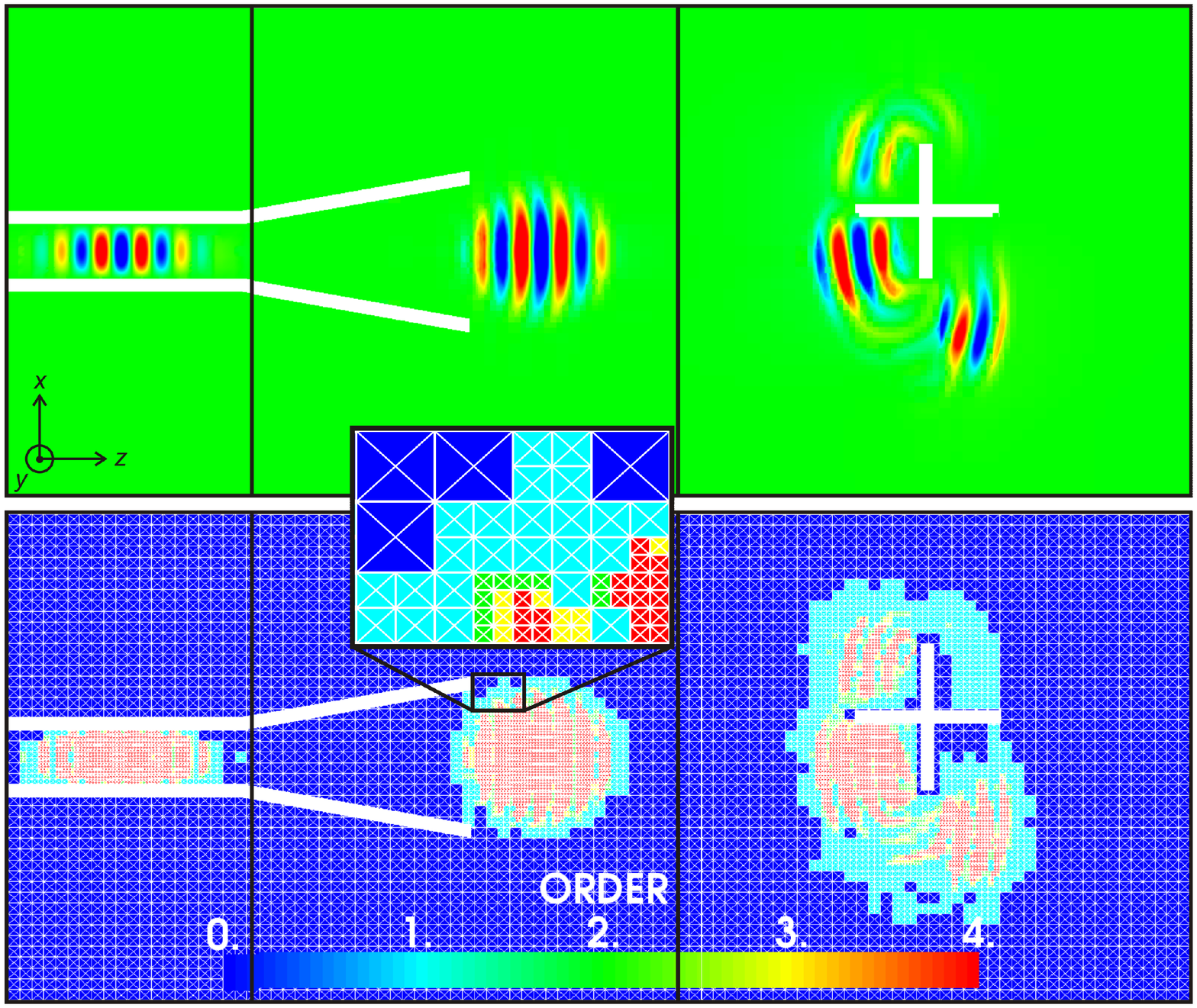}
  \caption{Visualizations of the $y$-component of the electric field (top panel) and
    the computational grid (bottom panel) at three instances in time. The enlargement shows details
    of the grid. We employ hexahedral elements for the computation but make use of
    embedded tetrahedra for displaying the tensor product orders in the grid view.
    As isotropic $p$-refinement was employed in this examples all tetrahedra
    associated with one element share a common color.
    Note that different scalings are used for the time instances in the top panel.}
  \label{fig:eyGrids}
\end{figure}

\begin{figure}[tbhp]
  \centering
  \includegraphics[width=0.75\textwidth]{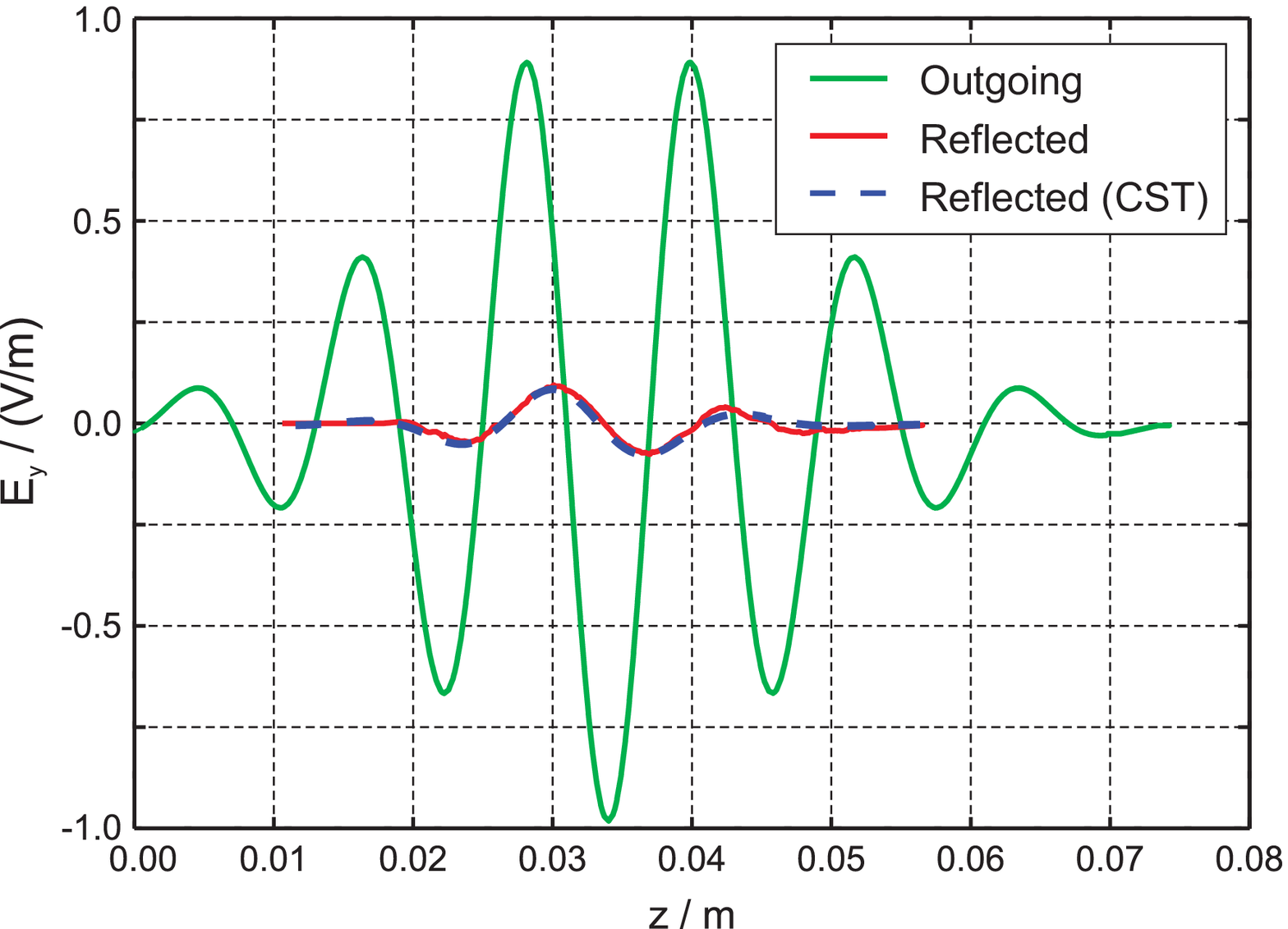}
  \caption{Plots of the outgoing and reflected waveform along the waveguide center. For crosschecking the setup was
  simulated using the commercial CST Microwave Studio software~\cite{CST} as well. The results are in a good agreement.}
  \label{fig:signals}
\end{figure}

Figure~\ref{fig:eyGrids} depicts cut-views of the $y$-component of the electric field and the respective
$hp$-mesh at three instances in time. Note that the scaling of the electric field
differs for every time instance, which is necessary to allow for a visual inspection.
The enlargement shows details of the computational grid. All elements are of hexahedral
kind, however, we make use of the common tensor product visualization technique
(cf.~\cite{Solin2008117,Demkowicz1}) using embedded tetrahedra for displaying the three tensor product orders
(out of which only $P_x$ and $P_z$ are visible in the depicted $x-z$-plane). As
we employed isotropic $h$- as well as $p$-refinement all tetrahedra associated
with one element share the same color.
Figure~\ref{fig:signals} shows plots of the outgoing and reflected waveform
recorded along the waveguide center. The blue dashed line was obtained with
the commercial CST Microwave Studio software~\cite{CST} on a very fine mesh
and serves as a cross comparison result.

\section{Conclusions}
\label{sec:conclusions}

We presented a discontinuous Galerkin formulation for non-regular
hexahedral meshes and showed that hanging nodes of high level can
easily be included into the framework. In fact, any non-regularity
of the grid can be included in a single term reflecting the contribution
of neighboring elements to the local interface flux. We demonstrated
that the method can be implemented such that it maintains its computational
efficiency also on non-regular and locally refined meshes as long as
the mesh is derived from a regular root tesselation by means of 
element bisections. This is achieved by extensive tabulations
of flux and projection matrices, which are obtained through (analytical) precomputations
of integral terms.

We also presented local refinement techniques for $h$- and $p$-refinements,
which are based on projections between finite element spaces. These
projections were shown to guarantee minimal projection errors in the $L^2$-sense
and to lead to an overall stable time-domain scheme.

Local error and smoothness estimates have been addressed, both of them
relate to the size of the interface jumps of the DG solution. We considered the simulation of
a smooth and a non-smooth waveform in a one-dimensional domain for
validating the error and smoothness estimates. 

As an application example in three-dimensional space the backscattering of a broadband waveform from
a radar reflector was considered. In this example the total wave propagation distance corresponds
to approximately sixty wavelengths involving thousand of local mesh adaptations. As the implementation of the derived error
and smoothness estimates for three-dimensional problems is subject of
ongoing work, we chose to drive the grid adaptation using the energy density as refinement indicator.
Crosschecking with a result obtained using a commercial software package showed
good agreement.












\end{document}